# Balancing Semantic Relevance and Engagement in Related Video Recommendations

Amit Jaspal
Meta Platforms, Inc.
Menlo Park, CA, USA
ajaspal@meta.com

Feng Zhang
Meta Platforms, Inc.
Menlo Park, CA, USA
fengzhang1@meta.com

Wei Chang
Meta Platforms, Inc
Menlo Park, CA, USA
mrweichang@meta.com

Sumit Kumar
Meta Platforms, Inc
Menlo Park, CA, USA
sumitkumar@meta.com

Yubo Wang
Meta Platforms, Inc
Menlo Park, CA, USA
yubowang@meta.com

Roni Mittleman
Meta Platforms, Inc
Menlo Park, CA, USA
romittel@meta.com

Qifan Wang
Meta Platforms, Inc
Menlo Park, CA, USA
wqfcr@meta.com

Weize Mao
Meta Platforms, Inc
Menlo Park, CA, USA
wzmao@meta.com

*Abstract*—Related video recommendations commonly use collaborative filtering (CF) driven by co-engagement signals, often resulting in recommendations lacking semantic coherence and exhibiting strong popularity bias. This paper introduces a novel multi-objective retrieval framework, enhancing standard two-tower models to explicitly balance semantic relevance and user engagement. Our approach uniquely combines: (a) multi-task learning (MTL) to jointly optimize co-engagement and semantic relevance, explicitly prioritizing topical coherence; (b) fusion of multimodal content features (textual and visual embeddings) for richer semantic understanding, and (c) off-policy correction (OPC) via inverse propensity weighting to effectively mitigate popularity bias. Evaluation on industrial-scale data and a two-week live A/B test reveals our framework's efficacy. We observed significant improvements in semantic relevance (from 51% to 63% topic match rate), reduction in popular item distribution (-13.8% popular video recommendations) and +0.04% improvement in our topline user engagement metric. Our method successfully achieves better semantic coherence, balanced engagement, and practical scalability for real-world deployment.

## I. INTRODUCTION

Recommending relevant subsequent videos is vital for user retention and content discovery on streaming platforms [1]. Related Item Recommendation (RIR) systems typically employ collaborative filtering (CF), particularly embedding-based two-tower models (Fig. 1) trained on co-engagement signals e.g., co-watches, co-clicks [2]. While effective, CF methods relying solely on behavioral signals face limitations: (a) Semantic ambiguity due to noisy co-engagements labels - Fast-paced video ecosystems produce superficial interactions e.g., autoplay, trending pages that can mislead models, associating semantically unrelated videos. (b) Pervasive popularity bias - CF inherently favors popular videos, creating feedback loops that marginalize niche content, reducing diversity and relevance. (c) Lack of interpretability: Purely behavior-based models lack transparency, diminishing user trust and complicates model debugging. To address these challenges, we propose a principled framework explicitly integrating semantic understanding into RIR systems.

Our contributions include:

- Multi-task learning (MTL), which combines co-engagement prediction with an auxiliary semantic alignment task, ensuring embeddings capture both behavior and semantic relationships.

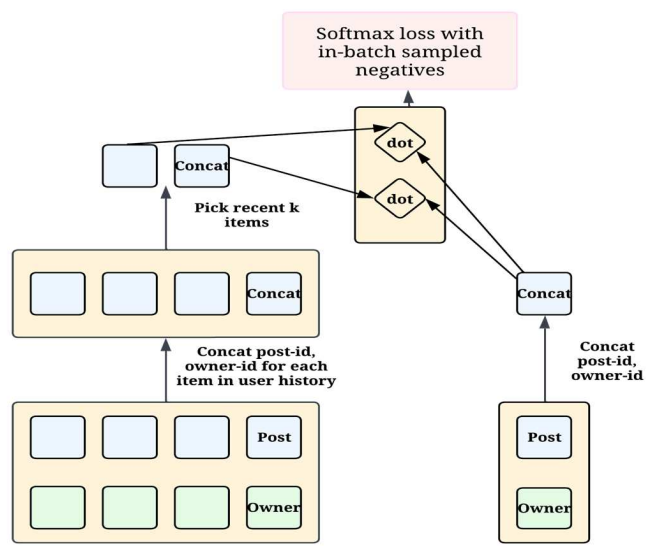

Fig. 1. Collaborative filtering with embedding based two tower model

- Fusion of multimodal content features, including textual metadata and visual content embeddings, enhancing semantic representation

- Off-policy correction (OPC) via inverse propensity weighting to mitigate popularity bias by down-weighting highly popular training samples.

To our knowledge, our work is the first to combine these 3 features in a single related-video recommendation model. This integrated approach is crucial to achieve the balanced outcomes we observe.

## II. PROPOSED APPROACH

Our proposed solution builds upon a conventional item-to-item retrieval framework (Fig. 1), integrating multi-task learning, multimodal content features, and bias mitigation tailored explicitly for multimedia recommendation scenarios.

## A. Multi-Task Learning for Co-Engagement and Semantic Relevance

We start with a two-tower neural embedding model, wherein two video inputs—a trigger video and a candidate video—are passed through shared-weight embedding networks to produce their latent embeddings. Traditionally, the similarity between embeddings indicates the likelihood of videos being co-engaged by users. We enhance this framework by jointly optimizing two complementary tasks (Fig. 2.):

### 1) User Co-engagement Task

We frame co-engagement prediction as a binary classification task $y_{tc}$. Given video pairs comprising a trigger video $v_t$ and a candidate video $v_c$, we define ground truth labels indicating co-engagement. The model predicts the probability of co-engagement using sampled softmax :

$$P(y_{tc} = 1 \mid v_t, v_c) = \frac{e^{v_t^T v_c}}{e^{v_t^T v_c} + \Sigma_{v_n \sim Q} e^{v_t^T v_n}} \quad (1)$$

Where $v_t^T v_c$ is the inner product between embeddings of the trigger video $v_t$ and candidate video $v_c$. $v_n \sim Q$ indicates random negative samples drawn from the same mini batch. The associated loss function is formulated as the negative log-likelihood:

$$L_{co-engage} = -\log\left(\frac{e^{v_t^T v_c}}{e^{v_t^T v_c} + \Sigma_{v_n \sim Q} e^{v_t^T v_n}}\right) \quad (2)$$

### 2) Semantic Similarity Aware Co-engagement Task :

To explicitly capture semantic relationships between co-engaged video pairs, we introduce an auxiliary task guided by pseudo-labels.

These labels are assigned based on whether the cosine similarity between BERT-encoded metadata representations of the two videos exceeds a predefined threshold. A binary indicator is then used to mark pairs that are both semantically similar and co-engaged :

$$I(sem_{t,c}) = \begin{cases} 1, & \text{if } v_t, v_c \text{ are semantically similar} \\ & \text{and co-engaged} \\ 0, & \text{otherwise} \end{cases} \quad (3)$$

We employ an additional semantic-aware sampled softmax loss to emphasize semantic coherence:

$$L_{semantic} = -I(sem_{t,c}) \cdot \log\left(\frac{e^{v_t^T v_c}}{e^{v_t^T v_c} + \Sigma_{v_n \sim Q} e^{v_t^T v_n}}\right) \quad (4)$$

Here:

- $v_t^T v_c$ denotes the inner product between trigger and candidate video embeddings.
- $v_n \sim Q$ are negative samples drawn from a negative sampling distribution $Q$.

The final training objective combines both losses as follows:

$$L = w_{sem} \cdot L_{semantic} + w_{co-engage} \cdot L_{co-engage} \quad (5)$$

In practice, we empirically set larger than (e.g., 10:1 or 100:1 ratio) to significantly emphasize semantic alignment in recommendations.

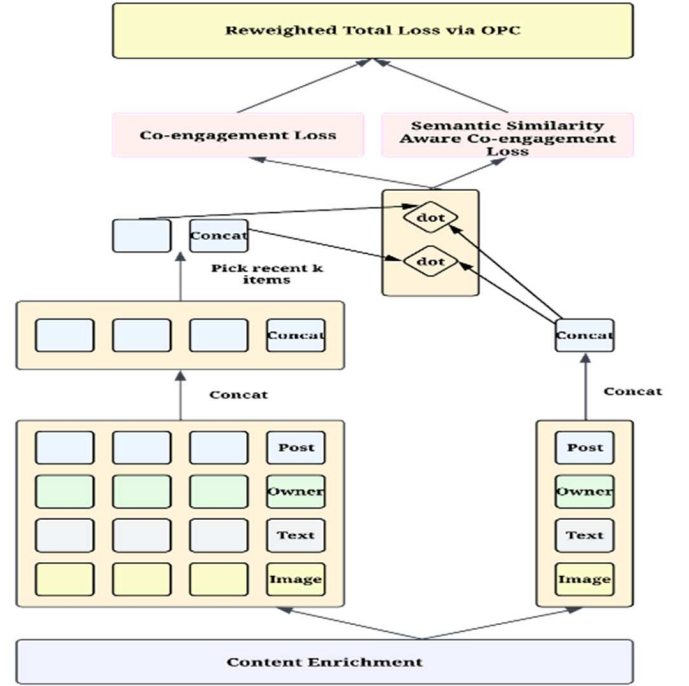

Fig. 2. Item-to-item collaborative filtering with multi-task learning, multimodal fusion and popularity bias mitigation

## B. Multimodal Content Integration for Semantic Representation

To robustly capture multimodal semantics, we incorporate metadata information into the representation of each video, extending beyond traditional collaborative filtering approaches:

- Textual Embeddings: Video titles, descriptions, and tags are encoded into semantic embeddings using transformer-based language models like BERT. These embeddings capture conceptual and contextual attributes of videos.

- Visual and Audio Embeddings: To directly leverage multimedia content, we extract embeddings from representative video frames using convolutional neural networks (CNNs). These embeddings encapsulate visual styles, object presence, scenes, and audio characteristics.

We combine these multimodal embeddings with collaborative embeddings via concatenation (Fig. 2), followed by a fully connected dense layer, yielding enriched final representations. Training gradients simultaneously refine collaborative and multimodal embeddings, enhancing the recommendation model's ability to identify semantically relevant videos.

## C. Off-Policy Correction (OPC) for Popularity Bias Mitigation

Despite leveraging multi-task learning and multimodal features, highly popular videos can still dominate the training distribution. For instance, a popular video may appear as the trigger in numerous training pairs, leading to overfitting and reduced semantic relevance. To address this, we introduce an Off-Policy Correction (OPC) mechanism inspired by reinforcement learning. OPC dynamically reweights each training instance based on trigger video popularity, reducing the influence of overrepresented content.

We define a dynamic weight $w_i$ for each training example $i$, inversely proportional to the frequency of the trigger video:

$$L_i = w_i \cdot (w_{sem} \cdot L_{semantic} + w_{co-engage} \cdot L_{co-engage}) \quad (6)$$

$$w_i = \frac{1}{\log(1 + freq(trigger_i))} \quad (7)$$

In this formulation, infrequent (niche) content receives a higher weight, allowing its co-engagement and semantic signals to significantly influence training. Conversely, frequently occurring (popular) content is assigned a much lower weight

## III. OFFLINE EXPERIMENTS

We evaluate our model on a large-scale video recommendation dataset derived from user sessions on our video platform. Positive co-engagement pairs are extracted from sequential views within or nearby sessions. Our experiments validate three hypotheses -- (H1) MTL improves semantic alignment without hurting engagement, (H2) Multimodal information boosts semantic relevance, and (H3) OPC improves semantic relevance. We measure engagement using standard recall@k [21], and semantic relevance using average topic overlap percentage between trigger and candidate video. We perform an ablation study comparing:

- Baseline CF: Standard item-CF on co-engagement only.
- MTL: Adds the semantic task (Tests H1).
- MTL + multimodal fusion : Adds content features (Tests H2).
- MTL + multimodal fusion + OPC: Full model with Off-Policy Correction (Tests H3).

### A. H1: Impact of Multi-Task Learning on Engagement vs. Semantic Similarity

To validate H1, we focused on the comparison between the Baseline CF, and MTL models. Specifically, we experimented with different weighting ratios (Table I) for the co-engagement loss weight $w_{co-engage}$ and the semantic similarity loss weight $w_{sem}$ within the MTL model.

TABLE I. ENGAGEMENT VS SEMANTIC SIMILARITY TRADEOFF WITH MULTI-TASK LEARNING

| $w_{co-engage} : w_{sem}$ | Engagement | Relevance |
|---|---|---|
| Baseline CF | 39.0% | 35.1% |
| Content Only Baseline | 15.5% | 71.2% |
| 1:1 | 38.6% | 39.5% |
| 1:10 | 38.4% | 41.2% |
| 1:100 | 38.1% | 45.1% |
| 1:500 | 37.9% | 46.9% |
| 1:1000 | 36.5% | 47.2% |

We also included content only baseline version with KNN directly with content embeddings. As detailed in Table I below, incorporating the semantic task consistently yields substantial improvements in semantic relevance metrics. Crucially, this significant boost in semantic alignment is achieved with only a minor reduction in engagement performance. These results confirm H1, demonstrating that MTL effectively enhances semantic coherence with a favorable trade-off against engagement prediction accuracy.

### B. H2: Enhancing Semantic Relevance and Engagement with multimodal features

Hypothesis H2 posits that incorporating multi-modal features leads to semantically more relevant recommendations. To test this, we compare the MTL model with most optimal tradeoffs (using the 1:500 weight ratio from H1) against variants that additionally incorporate item side features: text embeddings (from titles/descriptions), video frame embeddings, and a combination of both (MTL + Multimodal features).

Table II presents the results of this ablation study. The MTL model (without multimodal features) serves as the baseline for this comparison, achieving 37.9% engagement and 46.9% relevance. As shown in Table II, adding multimodal feature provides benefits on both fronts. Text embeddings significantly boost relevance to 50.5% (+3.6% absolute over MTL) and provide a slight lift in engagement to 38.1% (+0.2%). Similarly, video embeddings increase relevance to 49.0% (+2.1%) while marginally improving engagement (+0.1%).

Combining both text and video information in the MTL + multimodal features variant yields the best performance overall. It achieves the highest semantic relevance score of 52.0% (+5.1% absolute over MTL) and also delivers the highest engagement score of 38.3% (+0.4% absolute over MTL). This demonstrates that richer item representations derived from multimodal features not only improve semantic alignment but can also enhance the model's ability to predict user engagement, likely by capturing content nuances missed by collaborative signals alone.

TABLE II. ENHANCING SEMANTIC RELEVANCE WITH MULTIMODAL FUSION

| Multimodal information type | Engagement | Relevance |
|---|---|---|
| Baseline MTL | 37.9% | 46.9% |
| MTL + text embeddings | 38.1% | 50.5% |
| MTL + video embeddings | 38.0% | 49.0% |
| MTL + text and video embedding | 38.3% | 52.0% |
| Baseline CF (Single Task) | 39.0% | 35.1% |
| CF (Single Task) + text and video embedding | 39.2% | 37.8% |

### C. H3: Improving Relevance-Engagement Balance with Off-Policy Correction

Hypothesis H3 proposes that Off-Policy Correction (OPC) improves the balance between semantic relevance and engagement, particularly enhancing relevance for less popular items. To test this, we compare the MTL + multimodal features version with the full proposed model, MTL + multimodal features + OPC.

We analyze overall metrics and break down semantic relevance based on the popularity of the trigger video. Table III demonstrates the impact of OPC while there is a slight decrease in the engagement metric (-0.5% absolute), often expected as OPC corrects for exposure bias inherent in historical logs, there is a notable gain in overall semantic relevance (+2.5% absolute).

Breakdown by trigger popularity reveals OPC's effectiveness in mitigating bias. For popular trigger videos, relevance sees a minor decrease (-1.5%). However, for non-popular triggers, semantic relevance shows a substantial improvement (+7.2% absolute). These results validate H3. OPC improves the overall relevance-engagement profile by significantly boosting semantic relevance, especially for non-popular content where popularity bias is typically stronger. This leads to more balanced and semantically coherent recommendations across the item catalogue, with a small trade-off in the raw engagement metric.

### IV. ONLINE EXPERIMENTS

To evaluate the real-world performance of our proposed framework, we conducted a large-scale online A/B test over a two-week period on the video recommendation surface within our platform. The proposed model, integrating Multi-Task Learning (MTL), multimodal information fusion and Off-Policy Correction (OPC), was deployed against the existing production baseline system (primarily co-engagement based CF). The reported improvements in relevance and engagement metrics are statistically significant.

The online experiment revealed a subtle shift in engagement metrics. While short-term engagement metrics (proprietary) saw a slight decline—indicative of reduced popularity bias—we observed a +0.04% lift in a topline engagement metric (proprietary) tied to long-term user retention. This shift aligns with the model's design goals: a -13.8% drop in the distribution share of highly popular videos and a more balanced video length mix. Users engaged quickly with more semantically relevant content and continued to scroll for deeper discovery. Notably, semantic relevance—measured via trigger-candidate topic alignment—improved from 51% to 63%. These outcomes show that model effectively mitigates popularity bias while enhancing relevance, all within acceptable serving latency constraints.

TABLE III. ENHANCING SEMANTIC RELEVANCE WITH OFF POLICY CORRECTION

| Multimodal information type | Engagement | Relevance |
|---|---|---|
| Baseline(MTL + multimodal features) | 38.3% | 52% |
| Baseline + OPC | 37.8% | 54.5% |
| Baseline (Popular triggers, top 1%) | 50.5% | 61.2% |
| Baseline + OPC (Popular triggers, top 1%) | 49.4% | 59.7% |
| Baseline (Non popular triggers, bottom 10%) | 25.9% | 34.3% |
| Baseline + OPC (Non popular triggers, bottom 10%) | 32.4% | 41.5% |

### V. RELATED WORK

Related Item Recommendation (RIR) often leverages Collaborative Filtering (CF), particularly embedding-based two-tower models trained on co-engagement signals [2]. While effective at modeling user preferences, pure CF struggles with semantic relevance and suffers from popularity bias [3,4]. Hybrid approaches attempt to mitigate CF's limitations by incorporating content features (e.g., tags, metadata) or semantic information from external sources [5]. Multi-Task Learning (MTL) has been explored to balance multiple objectives, though often prioritizing engagement over explicit semantic alignment [6]. Bias mitigation techniques, including re-weighting and Off-Policy Correction (OPC) using inverse propensity scoring, are known strategies to counteract popularity bias [7]. Our work uniquely integrates these concepts for dynamic video RIR. We propose a unified framework featuring: (i) MTL explicitly balancing co-engagement and a primary semantic alignment objective, (ii) fusion of rich multimodal content features (textual/visual), and (iii) OPC tailored for mitigating popularity bias in this multi-objective setting. Unlike prior work that addresses these objectives in isolation, we present the first unified approach combining these features in a single related-video recommendation model. This integrated approach is crucial to achieve the balanced outcomes we observe.

REFERENCES

[1] C. A. Gomez-Uribe and N. Hunt, "The Netflix recommender system: Algorithms, business value, and innovation," ACM Trans. Manage. Inf. Syst., vol. 6, no. 4, Art. no. 13, Jan. 2016.

[2] G. Linden, B. Smith, and J. York, "Amazon.com recommendations: Item-to-item collaborative filtering," *IEEE Internet Comput.*, vol. 7, no. 1, pp. 76–80, Jan.–Feb. 2003.

[3] C. Desrosiers and G. Karypis, "A comprehensive survey of neighborhood-based recommendation methods," in Recommender Systems Handbook. Springer, 2011.

[4] B. M. Sarwar, G. Karypis, J. A. Konstan, and J. Riedl, "Item-based collaborative filtering recommendation algorithms," in Proc. 10th Int. Conf. World Wide Web (WWW '01), Hong Kong, May 1-5, 2001, pp. 285–295.

[5] Z. Xu, C. Chen, T. Lukasiewicz, Y. Miao, and X. Meng, "Tag-aware personalized recommendation using a deep-semantic similarity model with negative sampling," in Proc. 25th ACM Int. Conf. Inf. Knowl. Manage. (CIKM '16), Indianapolis, IN, USA, Oct. 24–28, 2016, pp. 1943–1946..

[6] T. Bansal, D. Belanger, and A. McCallum, "Ask the GRU: Multi-task learning for deep text recommendations," in *Proc. 10th ACM Conf. Recommender Syst. (RecSys '16)*, Boston, MA, USA, Sep. 2016, pp. 207-214

[7] D. Liang, L. Charlin, J. McInerney, and D. M. Blei, "Modeling user exposure in recommendation," in *Proc. 25th Int. Conf. World Wide Web (WWW '16)*, 2016, pp. 135-144